\documentclass[reprint,nofootinbib,amsmath,amssymb,aps,floatfix]{revtex4-2}
\usepackage{graphicx,hyperref}
\usepackage{dcolumn}
\usepackage[multiple]{footmisc}
\usepackage{appendix}
\usepackage{color}
\usepackage{multirow}
\def\a{\alpha}\def\b{\beta}

\newcommand{\nn}{\nonumber}

\begin{document}
\title{Quasinormal modes of massive scalar fields in five-dimensional Myers-Perry black holes with two arbitrary rotation parameters}
\author{Zi-Yang Huang}
\author{Jia-Hui Huang} 
\email{huangjh@m.scnu.edu.cn}
\date{\today}

\affiliation{Key Laboratory of Atomic and Subatomic Structure and Quantum Control (Ministry of Education), Guangdong Basic Research Center of Excellence for Structure and Fundamental Interactions of Matter, School of Physics, South China Normal University, Guangzhou 510006, China}
\affiliation{Guangdong Provincial Key Laboratory of Quantum Engineering and Quantum Materials, Guangdong-Hong Kong Joint Laboratory of Quantum Matter, South China Normal University, Guangzhou 510006, China}

\begin{abstract}
We investigate the quasinormal modes of massive scalar fields in the background of five-dimensional Myers-Perry black holes. In particular, we explore the case for Myers-Perry black holes with two arbitrary rotation parameters. Since the Klein-Gordon equation for the scalar field is separable, we numerically compute the scalar quasinormal modes by using the radial and angular equations. Two methods, the continued fraction method and matrix method, are used in the numerical calculation. We find that all obtained modes have negative imaginary parts and are decaying modes. We also consider the impact of the rotation parameters, scalar field mass $\mu$ and azimuthal numbers on the scalar quasinormal modes. 
Besides, when the scalar mass $\mu$ becomes relatively large, we also find the long-living scalar modes. Our numerical results also demonstrate the symmetries of the QNMs explicitly.   
\end{abstract}
\maketitle

\section{Introduction}
Black holes play important roles in modern theoretical and observational physics. An important issue about black hole solutions is their (in)stability under various perturbations. When a black hole is perturbed by an external field (e.g. scalar, electromagnetic, gravitational), the evolution of the perturbation field is usually governed by some second-order differential equations. With appropriate boundary conditions, we can obtain various characteristic modes of the perturbation field. The linear (in)stability issue can be addressed by analyzing the perturbation modes. The quasinormal modes (QNMs) discussed in this work are determined by requiring purely ingoing wave at the event horizon and purely outgoing wave at spatial infinity \cite{Berti:2009kk,Konoplya:2011qq}.

Four-dimensional Kerr black hole is stable under massless scalar, electromagnetic or gravitational perturbations \cite{Press:1973zz,Teukolsky:1974yv}. However, a massive scalar field may lead to instability due to the mass term of the perturbation field \cite{Damour:1976jd, Detweiler:1980uk,Furuhashi:2004jk,Dolan:2007mj}. The superradiant (in)stability of asymptotically flat Kerr black holes under massive scalar and vector perturbation has been studied extensively in the literature \cite{Strafuss:2004qc,Konoplya:2006br,Cardoso:2011xi,Dolan:2012yt,Hod:2012zza,Hod:2014pza, Aliev:2014aba,Hod:2016iri,Degollado:2018ypf,Xu:2020fgq,Huang:2019xbu,Lin:2021ssw,East:2017ovw,East:2017mrj}.

Higher-dimensional spacetime is interesting in theoretical physics. On the one hand, higher dimensions are necessary for string theory, brane-world models, gauge/gravity duality, etc. On the other hand, there are a variety of black object solutions in higher-dimensional spacetime \cite{Emparan:2008eg}. 
The higher-dimensional extensions of Kerr black holes are Myers-Perry black holes (MPBHs) \cite{Emparan:2008eg, Myers:1986un}, and the extensions with nonvanishing cosmological constants are MP-(A)dS black holes \cite{Gibbons:2004js}.

The linear (in)stability of higher-dimensional black holes has been studied extensively in literature \cite{Brito:2015oca}. For example, the higher-dimensional Schwarzschild-Tangherlini black holes and Reissner-Nordstr\"{o}m  black holes were proved stable under perturbation of external fields \cite{Ishibashi:2003ap,Konoplya:2007jv,Konoplya:2013sba,Zhidenko:2006rs,Konoplya:2004wg,
Kodama:2003kk,Ishibashi:2011ws,Huang:2021jaz,huang2022-6d,huang2022-d,Huang:2021dpa,Huang:2023kzv}. 
The linear (in)stability of higher-dimensional rotating black holes is more complicated.  
Singly rotating MP-AdS black holes are unstable against tensor-type gravitational perturbation, and growing QNMs appear when the rotation parameter is larger than a critical value \cite{Kodama:2007sf, Kodama:2009rq}.
MP-AdS black holes with equal rotation parameters in odd number of dimensions (greater than five) are unstable under gravitational perturbations \cite{Kunduri:2006qa}.
Five-dimensional ($5D$) MP-AdS black holes with equal rotation parameters were found unstable against gravitational perturbation when the black holes are ultraspinning \cite{Cardoso:2013pza}. Small singly rotating MP-AdS black holes in arbitrary dimensions have also been proven unstable under massive scalar perturbations \cite{Delice:2015zga}. 
Analytical and numerical results of the scalar QNMs of $5D$ MP-AdS black holes with arbitrary rotation parameters were obtained in \cite{Amado:2017kao,BarraganAmado:2018zpa, Koga:2022vun}.
The QNMs of higher-dimensional singly rotating MP-dS black holes with non-minimally coupled scalar fields were analytically studied, and a formula for QNMs was derived \cite{Gwak:2019ttv}, and it is found that singly rotating MP-dS black hole is stable under a massive scalar perturbation \cite{Ponglertsakul:2020ufm}. Recently, the massive scalar QNMs of MP-dS with equal rotation parameters have also been studied \cite{Davey:2022vyx}.

For asymptotically flat MPBHs, singly rotating MPBHs against tensor-type perturbations were studied and no instability was found when $D \geq 7$ \cite{Kodama:2009bf}. 
MPBHs with equal rotation parameters were found to be stable under gravitational perturbation in five or seven dimensions, but in nine dimensions, for sufficiently rapid rotation, the perturbations grow exponentially in time \cite{Dias:2010eu,Murata:2008yx}. A thorough study of the QNMs of the asymptotically flat singly rotating MPBHs and odd-dimensional MPBHs with equal rotation parameters were presented in \cite{Dias:2014eua}. The large $D$ limit analysis of the QNMs of rotating black holes was discussed in \cite{Emparan:2014jca,Emparan:2015rva}.

The QNMs of massless scalar perturbations on $5D$ and $6D$ singly rotating MPBHs were well investigated, and the numerical results show that they are all stable \cite{Ida:2002zk,Cardoso:2004cj,Morisawa:2004fs}. It was argued that singly rotating MPBHs should be stable against massive scalar perturbations \cite{Cardoso:2005vk}. Recently, we have explicitly calculated the QNMs of massive scalar perturbations on $5D$ and $6D$ singly rotating MPBHs and no unstable modes were found \cite{Lu:2023par,Li:2023wog}.
In this work, we go a step further and consider the massive scalar QNMs of $5D$ MPBHs with two arbitrary rotation parameters.

This paper is organized as follows. In Sec.II, we briefly review the Klein-Gordon equation in $5D$ MPBH background, and present the radial and angular equations of motion (EOMs). 
In Sec.III, we give brief introduction to the numerical methods and calculate the massive scalar QNMs. The impact of model parameters on the massive scalar QNMs are also discussed.  
The final section is devoted to the summary.

\section{Myers-Perry black holes and equations of motion}

In this work, we concentrate on $5D$ MPBHs with two arbitrary rotation parameters denoted by $a$ and $b$. In Boyer-Lindquist-type coordinates, the line element of such a MPBH can be written as \cite{Myers:1986un}

\begin{align}
     \mathrm{d}s^2 = & - \left(1-\frac{M}{\rho^2}\right) \mathrm{d} t^2 + \frac{r^2 \rho^2}{\Delta} \mathrm{d} r^2+\rho^2 \mathrm{d}\theta^2 \nn\\
     &- \frac{2 a M \sin^2 \theta}{\rho^2} \mathrm{d}t \mathrm{d}\phi  - \frac{2 b M \cos^2 \theta}{\rho^2} \mathrm{d}t \mathrm{d}\psi \nn\\
     &+ \frac{2 a b M \sin^2 \theta \cos^2 \theta}{\rho^2} \mathrm{d}\phi \mathrm{d}\psi \nn\\
    &+ \sin^2 \theta \left(r^2 + a^2 + \frac{M a^2 \sin^2 \theta}{\rho^2}\right) \mathrm{d}\phi^2 \nn\\
    &+ \cos^2 \theta \left(r^2 + b^2 + \frac{M b^2 \cos^2 \theta}{\rho^2}\right) \mathrm{d}\psi^2 ,\label{2.1}
\end{align}
where
\begin{align}
	\rho^2 &= r^2 + a^2 \cos^2 \theta+ b^2 \sin^2 \theta, \label{2.2} \\
	\Delta &= \left(r^2 + a^2\right) \left(r^2 + b^2\right)- M r^2. \label{2.3}
\end{align}
 $M$ is proportional to the Arnowitt-Deser-Misner mass of the black hole, and $a,b,$ are proportional to the angular momenta \cite{An:2017phb}. The coordinate ranges are $0\leq\theta\leq\pi/2, 0\leq\phi,\psi\leq2\pi$. 
 The metric\eqref{2.1} describes an asymptotically flat and rotating vacuum BH solution with spherical topology. Hereafter, without loss of generality,  $M,a$ and $b > 0$ are assumed.

The location of the event horizon of the MPBH is the biggest real root of equation $\Delta=0$. When $M>(a+b)^2$, there are totally four real roots, which are
\begin{align}
      r_{1,4}=&\pm\frac{\sqrt{\left(M - a^2 - b^2\right) + \sqrt{(M - a^2 - b^2)^2 - 4a^2b^2} }}{\sqrt{2}},\nn\\
      r_{2,3}=&\pm\frac{\sqrt{\left(M - a^2 - b^2\right) - \sqrt{(M - a^2 - b^2)^2 - 4a^2b^2} }}{\sqrt{2}}.\nn
\end{align}
The event horizon is at $r_1(>0)$ and inner horizon is at $r_2(>0)$. The MPBH becomes degenerate if $a=b$ \cite{Murata:2007gv}.  More discussion of $M$ and $(a+b)^2$ can be found in \cite{An:2017phb} .  

The dynamics of a scalar perturbation $\Psi(x)$ with mass $\mu$ on $5D$ MPBH is governed by the covariant Klein-Gordon equation
\begin{equation}\label{2.5}
	\Box \Psi(x) = \dfrac{1}{\sqrt{-g}} \partial_\a\left[ g^{\a \b} \sqrt{-g} \
 \partial_\b\Psi(x) \right]  = \mu^2 \Psi(x), 
\end{equation}
where $g=\det(g_{\a \b})$ is the determinant of the metric. The above equation is separable in the Boyer-Lindquist-type coordinates $x^\mu = \left\lbrace t,r,\theta,\varphi,\psi\right\rbrace $ \cite{Frolov:2002xf}. We choose the following ansatz for the scalar field 
\begin{equation}\label{2.6}
	\Psi(x) = e^{-i \omega t} e^{i m_1 \varphi} e^{i m_2 \psi} R(r) S(\theta), 
\end{equation}
where $\omega$ is the angular frequency and $m_1, m_2 \in \mathbb{Z}$ are azimuthal numbers. Plugging the above ansatz into Eq.\eqref{2.5}, we obtain the radial and angular EOMs.
The angular EOM obeyed by $S(\theta)$ is
\begin{align}
	\dfrac{1}{\sin \theta \cos \theta} & \dfrac{\mathrm{d}}{\mathrm{d}\theta} \left(\sin \theta \cos \theta \dfrac{\mathrm{d} S(\theta)}{\mathrm{d}\theta}\right)\nn\\
    &+ \bigg[ \left(\omega^2 - \mu^2\right) \left(a^2 - b^2\right) \cos^2\theta\nn\\
    & - \frac{m_1^2}{\sin^2\theta} - \frac{m_2^2}{\cos^2\theta} +\lambda_{\substack{km_1 m_2}} \bigg] S(\theta) = 0.\label{2.7}
\end{align}
The above equation is the $5D$ spheroidal equation of which the solutions are scalar spheroidal harmonics \cite{Frolov:2002xf,Berti:2005gp}. Since the above equation is invariant under $m_1\rightarrow -m_1, m_2\rightarrow -m_2$, we restrict our consideration to $m_1\geq0, m_2\geq0$.
$\lambda_{\substack{km_1 m_2}}$ is the separation constant and $k=\left\lbrace0,1,2,\cdots\right\rbrace$ labels the discrete eigenvalues of $S(\theta)$ for given $m_1$ and $m_2$.
When $|\left(\omega^2 - \mu^2\right) \left(a^2 - b^2\right)| \ll 1$ , $\lambda_{\substack{km_1 m_2}}$ can be expanded as a series
\begin{equation}\label{2.8}
	\lambda_{\substack{km_1 m_2}} = \ell (\ell+ 2) + \sum_{p = 1}^{\infty} \tilde{f}_p c^p,
\end{equation}
where $\ell= 2k + m_1 + m_2$  and  $c=\sqrt{\left(\omega^2 - \mu^2\right) \left(a^2 - b^2\right)}$ .  $\ell$ is an integer and  $\ell \geq m_1 + m_2$. More details of  $\tilde{f}_p$ can be found in \cite{Berti:2005gp}.

The radial EOM obeyed by $R(r)$ is
\begin{equation}\label{3.1}
	\dfrac{1}{r} \dfrac{\mathrm{d}}{\mathrm{d}{r}} \left(  \dfrac{\Delta}{r}  \dfrac{\mathrm{d} R(r)}{\mathrm{d} r} \right) + U(r) R(r) = 0,
\end{equation}
where
\begin{widetext}
	\begin{equation}\label{3.2}
		\begin{aligned}
			U(r) = &\left[ \frac{(b^2 + r^2)^2 a^2}{\Delta r^2} - \frac{b^2}{r^2} \right] m_1^2
 + \left[ \frac{(a^2 + r^2)^2 b^2}{\Delta r^2} - \frac{a^2}{r^2} \right] m_2^2 
- \frac{2 M r^2 a m_1 \omega}{\Delta} \left( 1 + \frac{b^2}{r^2} \right)  \\&
- \frac{2 M r^2 b m_2 \omega}{\Delta} \left( 1 + \frac{a^2}{r^2} \right) 
 + \frac{2 M a b m_1 m_2}{\Delta} 
+ \left( \frac{M}{r^2} + \frac{M^2}{\Delta} \right) r^2 \omega^2 
+ (r^2 + b^2) \left( \omega^2 - \mu^2 \right) - \lambda_{\substack{km_1 m_2}}. 
		\end{aligned}
	\end{equation}
\end{widetext}
It's easy to check that the above radial EOM reduces to that of a singly rotating $5D$ MPBH case when $b=0$ \cite{Li:2023wog}. 

Next we study the asymptotical behaviors of the scalar perturbation near boundaries. It is useful to define the tortoise coordinate $r_*$ by $\dfrac{\mathrm{d} r_*}{\mathrm{d} r} = \dfrac{\left(r^2 + a^2\right)\left(r^2 + b^2\right)}{\Delta}$ and a new radial function $\tilde{R}(r) = \sqrt{\dfrac{ (r^2 + a^2)(r^2 + b^2)}{r}} R(r)$. The radial EOM can be rewritten as a Schr\"{o}dinger-like equation,
\begin{equation}\label{3.3}
	\dfrac{\mathrm{d}^2 \tilde{R}(r)}{\mathrm{d} r_*^2} + \tilde U(r) \tilde{R}(r) = 0.
\end{equation}
Then, the limits of $\tilde U(r)$ at the event horizon and spatial infinity are
\begin{align}\label{3.4}
	\tilde U(r) \sim 
	\begin{cases}
		\left( \omega - m_1 \Omega_{\substack{H a}} - m_2 \Omega_{\substack{H b}} \right)^2, \quad {} \, r_* \rightarrow -\infty \, (r \rightarrow r_H)\\\\
		\omega^2 - \mu^2,  \quad {} \, r_* \rightarrow +\infty \, (r \rightarrow +\infty),
	\end{cases}
\end{align}
where $\Omega_{\substack{H a}} = \dfrac{a}{r_H^2 + a^2}$ , $\Omega_{\substack{H b}} = \dfrac{b}{r_H^2 + b^2}$, which can be easily found by using the trick proposed in\cite{Yang:2023kgc}. With the QNM boundary conditions, the asymptotic solutions of the radial equation \eqref{3.3} at the event horizon and spatial infinity are 
\begin{equation}\label{3.5}
	R(r) \sim 
	\begin{cases}
		(r-r_H)^{- i \sigma}, & \quad  r \rightarrow r_H, \\
		r^{-3/2} {\rm e}^{q r}, & \quad  r \rightarrow \infty,
	\end{cases}
\end{equation}
where
\begin{align}
	&q^2 = \mu^2-\omega^2, \nn\\\sigma =&\left( \omega -\dfrac{ m_1 a}{r_H^2+a^2} - \dfrac{ m_2 b}{r_H^2+b^2}\right)\times\nn\\& \dfrac{ \left( r_H^2 + a^2 \right) \left( r_H^2 + b^2 \right)}{2 r_H\left(a^2+b^2-M+2 r_H^2\right)}.
\end{align}
At spatial infinity, the QNM condition imposes purely outgoing waves at $r \rightarrow \infty$. In this case, 
${\rm Re}(\omega)>\mu$ and $q=i\sqrt{\omega^2-\mu^2}$. 
The angular and radial EOMs and the chosen boundary conditions single out a discrete set of complex frequencies $\left\lbrace \omega \right\rbrace$ ($\omega \equiv \omega_R + i \omega_I$). In our convention, $\omega_I< 0$ implies a stable mode.

\section{Numerical calculation of QNMs}

In this section, we numerically calculate the massive scalar QNMs. In the calculation, we use the continued fraction method \cite{Leaver:1985ax,Leaver:1990zz} and matrix method \cite{Lin:2016sch,Lin:2019mmf}.
We also take $M = 1$ for convenience, so $r$ and $a,b$ are scaled in unit of $M^{1/2}$, while $\omega$ and $\mu$ are scaled in $M^{-1/2}$. 

\subsection{Angular Equation of Motion}

Define a new variable, $u \equiv \cos \theta$, the angular EOM \eqref{2.7} can be rewritten as
\begin{align}
	\dfrac{1}{u} \bigg(\frac{\text{d}}{\text{d} u} u \left( 1 - u^2 \right)& \dfrac{\text{d}}{\text{d} u} \bigg)S(u) \nn 
	+ \bigg[ \left(\omega^2 - \mu^2\right) \left(a^2 - b^2\right) u^2 \\& - \dfrac{m_1^2}{1-u^2} - \dfrac{m_2^2}{u^2}+\lambda_{\substack{km_1 m_2}}  \bigg] S(u) = 0.\label{5.1}
\end{align}
The angular function $S(u)$ can be assumed to have the following expansion \cite{Berti:2005gp}
\begin{equation}\label{5.2}
	S = (1 - u^2)^{\frac{m_1}{2}} u^{m_2} \sum_{p = 0}^{\infty} a_p u^{2 p}.
\end{equation}
Substitute this expansion into Eq.\eqref{5.1}, we obtain a 3-term recursion relation 
\begin{align}
	& \alpha_0^\theta a_1 + \beta_0^\theta a_0 =0, \nonumber \\
	& \alpha_p^\theta a_{ p+1 } + \beta_p^\theta a_p + \gamma_p^\theta a_{ p-1 } =0 \qquad ( p = 1,2, 3, \cdots),
\end{align}
where
\begin{align}
	\alpha_p^\theta &= -4 (p+1) (m_2 + p + 1),\nn \\
	\beta_p^\theta &= \left( 2p + m_1 + m_2 \right) \left( 2p + m_1 + m_2 + 2 \right) - \lambda_{\substack{km_1 m_2}},\nn \\
	\gamma_p^\theta &= -\left(\omega^2 - \mu^2\right) \left(a^2 - b^2\right).
\end{align}
Then the continued fraction equation to determine the separation constant $ \lambda_{\substack{km_1 m_2}}$ and the QNM frequency $\omega$ is \cite{Leaver:1985ax}
\begin{equation}\label{5.5}
	\beta_0^\theta - \frac{ \alpha_0^\theta \gamma_1^\theta}{ \beta_1^\theta - } \frac{ \alpha_1^\theta \gamma_2^\theta }{ \beta_2^\theta - } \frac{ \alpha_2^\theta \gamma_3^\theta }{ \beta_3^\theta -  } \cdots 
	\equiv
	\beta_0^\theta - \frac{\alpha_0^\theta \gamma_1^\theta}{\beta_1^\theta - \frac{\alpha_1^\theta \gamma_2^\theta} {\beta_2^\theta - \frac{\alpha_2^\theta \gamma_3^\theta}
			{\beta_3^\theta - \cdots}}}
	= 0.
\end{equation}
When $a=b=0$, all $\gamma_p^\theta$ will be zero and the recursion will stop whenever $\lambda_{\substack{km_1 m_2}}$ is such that $\beta_p^\theta$ is zero for some $p$.
In this case, $\lambda_{\substack{km_1 m_2}}$ can be read from Eq.\eqref{2.8}, 
\begin{align}
	\lambda_{\substack{km_1 m_2}} =& \left( 2 k + m_1 + m_2 \right) \left( 2k + m_1 + m_2 + 2 \right), \nn \\
	&k=\left\lbrace0,1,2,\cdots\right\rbrace.\label{5.6}
\end{align}

\subsection{Radial Equation of Motion}

The radial EOM \eqref{3.1} besides $\infty$ has four singular points $r=r_{1,4},r_{2,3}$. This leads to more subtlety than the singly rotating MPBH case
when we use the continued fraction method to analyze this equation. And, we need a small inner horizon ($r_2$) approximation. Approximations at different orders result in recursion relations with different terms. For example, approximations at $0$-th and $1$-st orders result in a 12-term and 25-term recursion relations respectively. 
In order to calculate the QNMs when $r_2$ is not too small and to provide crosschecks between different methods, we here use the matrix method \cite{Lin:2016sch,Lin:2019mmf}. In the following subsections, we briefly introduce these two methods.

\subsubsection{Continued Fraction Method}

The solution of the radial EOM, Eq.\eqref{3.1}, can be assumed to have the following form 
\begin{align}
	R =& \left( \dfrac{r - r_1}{r - r_2} \right)^{ - i \sigma} \left( \dfrac{r - r_3}{r - r_4} \right)^{i \sigma - \frac{3}{2}} {\rm e}^{q r} \nn \\
	&\qquad \times \sum_{ p=0 }^{ \infty } b_p \left( \dfrac{r - r_1}{r - r_2} \times\dfrac{r - r_3}{r - r_4}\right)^p.\label{5.7}
\end{align}
Plugging the above function into Eq.\eqref{3.1}, we obtain a 12-term recursion relation for the expansion coefficients $\{b_p\}$ when $r_2\ll1$. 
The recursion equations have the following general forms, 
	\begin{align}
		\mathcal{A\substack{1}}_0^r b_1 + \mathcal{A\substack{2}}_0^r b_0 &= 0,\nn  \\
		\mathcal{A\substack{1}}_1^r b_2 + \mathcal{A\substack{2}}_1^r b_1 + \mathcal{A\substack{3}}_1^r b_0 &= 0,\nn  \\
		\mathcal{A\substack{1}}_2^r b_3 + \mathcal{A\substack{2}}_2^r b_2 + \mathcal{A\substack{3}}_2^r b_1 + \mathcal{A\substack{4}}_2^r b_0 &= 0,\nn\\
      \vdots \qquad \qquad\nn&\\
      \mathcal{A\substack{1}}_{n-3}^r b_{n-2} +\underbrace{\cdots}_{\text{($n-3$) terms}}+\mathcal{A\substack{n-1}}_{n-3}^r b_0 &= 0,\nn\\ 
		\mathcal{A\substack{1}}_p^r b_{p+1} + \mathcal{A\substack{2}}_p^r b_{p} + \mathcal{A\substack{3}}_p^r b_{p-1} + \mathcal{A\substack{4}}_p^r b_{p-2}\nn\\
 + \mathcal{A\substack{5}}_p^r b_{p-3} + \cdots+ \mathcal{A\substack{n}}_p^r b_{p-n+2}  &= 0,\nn\\(p= n-1,n,\cdots),  \label{5.13}
	\end{align}
where $\mathcal{A\substack{n}}_p^r$ depends on $\lambda_{km_1m_2}$, $\omega$ and other model parameters.
In our case, $n=12$ at 0-th approximation and $n=25$ at 1-st approximation. The explicit expressions of these coefficients are rather complicated and we will not show them here. 
Using Gaussian elimination, the recursion relation \eqref{5.13} can be reduced to a 3-term recursion relation
\begin{align}
&\tilde\alpha_0^r b_{ 1} + \tilde\beta_0^r b_0 = 0,\nn\\
&\tilde\alpha_p^r b_{p + 1} + \tilde\beta_p^r b_p + \tilde\gamma_p^r b_{p - 1} = 0 \qquad (p =1, 2,3,\cdots).
\end{align}
The expressions of $\tilde\alpha_p^r,\tilde\beta_p^r$ are rather complicated and will not be shown here. The continued fraction equation for the radial EOM is as follows,
\begin{equation}\label{5.15}
	\tilde\beta_0^r - {\tilde\alpha_0^r \tilde\gamma_1^r \over \tilde\beta_1^r - }
	{\tilde\alpha_1^r \tilde\gamma_2^r \over \tilde\beta_2^r - }
	{\tilde\alpha_2^r \tilde\gamma_3^r \over \tilde\beta_3^r - } \cdots = 0.
\end{equation}

For a given set of values of the parameters $\{k,m_1,m_2,a,b,\mu\}$, the QNM frequency $\omega$ and the separation constant $\lambda_{km_1m_2}$ can be obtained by solving the two coupled algebraic equations \eqref{5.5} and \eqref{5.15} simultaneously.

\begin{table}[tbph]
	\centering
	\caption{Numerical results of the inner horizon $r_2$. $M=1$.}
	\label{tab:2}
	\renewcommand\arraystretch{1.25}
	\setlength{\tabcolsep}{2.5pt}{
		\begin{tabular}{c| c |c |c |c| c| c| c}
			\hline
			\hline
			$b \backslash a$ & $0.1$ & $0.2$ & $0.3$ & $0.4$ & $0.5$ & $0.6$ \\
			\hline
			$0.1$ & $0.01010$ & $0.02052$ & $0.03164$ & $0.04396$ & $0.05826$ & $0.07594$ \\
            \hline
			$0.2$ & $0.02052$ & $0.04174$ & $0.06448$ & $0.089890$ & $0.11990$ & $0.15826$ \\
            \hline
			$0.3$ & $0.03164$ & $0.06448$ & $0.1$ & $0.14042$ & $0.18990$ & $0.25903$ \\
			\hline
            $0.4$ & $0.04396$ & $0.08990$ & $0.14042$ & $0.2$ & $0.27955$ & $-$ \\
			\hline
            $0.5$ & $0.05826$ & $0.11990$ & $0.18990$ & $0.27955$ & $-$ & $-$ \\
            \hline
            $0.6$ & $0.07594$ & $0.15826$ & $0.25903$ & $-$ & $-$ & $-$ \\
            \hline
             \hline
	\end{tabular}}
\end{table}

\subsubsection{Matrix Method}
By considering the QNM boundary conditions, we define a new radial function $\mathcal{Y}(r)$ by following equation
\begin{align}
	R =& \left( \dfrac{r - r_1}{r - r_2} \right)^{-i \sigma} \left( \dfrac{r - r_3}{r - r_4} \right)^{i \sigma - \frac{3}{2}} {\rm e}^{q r} \mathcal{Y}(r).\label{4.1}
\end{align}
$\mathcal{Y}$ is finite at the boundaries. Then introducing the following transformations of the radial coordinate and function
\begin{align}
	r\to&\dfrac{r_1}{1-x},\quad
    \mathcal{Y}\to\dfrac{\mathcal{Z}(x)}{(x-1)x},\label{4.2}
\end{align}
where $0\leq x\leq1$, and substituting \eqref{4.1}\eqref{4.2} into the radial EOM \eqref{3.1},
we obtain a differential equation for $\mathcal{Z}(x)$, 
\begin{align}
	\mathcal{C\substack{2}}(x)\mathcal{Z}''(x)+\mathcal{C\substack{1}}(x)\mathcal{Z}'(x)+\mathcal{C\substack{0}}(x)\mathcal{Z}'(x)=0.\label{4.3}
\end{align}
$\mathcal{Z}(x)$ satisfies the following boundary conditions
\begin{align}
\mathcal{Z}(0)&=\mathcal{Z}(1)=0.\label{4.03}
\end{align}

We expand the function $\mathcal{Z}(x)$ at $N$ discretized points $\lbrace{x_1,x_2,x_3,\cdots,x_N}\rbrace$. For example, when we expand it at point $x_j$, we have the following Taylor series
\begin{align}
	&\mathcal{Z}(x_i)=\mathcal{Z}(x_j) + (x_i -x_j)\mathcal{Z}'(x_j)\nn\\&+\frac{(x_i - x_j)^2}{2!} \mathcal{Z}''(x_j) + \frac{(x_i -x_j)^3}{3!} \mathcal{Z}'''(x_j) + \cdots\label{4.4}
\end{align}
for the function value at $x_i$ $(i=1,2,\cdots,i-1,i+1,\cdots,N)$. We approximate the expansions up to $(N-1)$-th order, then we can obtain following equations
\begin{align}
	\Delta F = M D,\label{4.5}
\end{align}
where
\begin{widetext}
\begin{align}
	M = \left(
\begin{array}{cccccc}
x_1 - x_j & \frac{(x_1 - x_j)^2}{2} & \cdots & \frac{(x_1 - x_j)^i}{i!} & \cdots & \frac{(x_1 - x_j)^{N-1}}{(N-1)!} \\
x_2 - x_j & \frac{(x_2 - x_j)^2}{2} & \cdots & \frac{(x_2 - x_j)^i}{i!} & \cdots & \frac{(x_2 - x_j)^{N-1}}{(N-1)!} \\
\cdots & \cdots & \cdots & \cdots & \cdots & \cdots \\
x_{j-1} - x_j & \frac{(x_{j-1} - x_j)^2}{2} & \cdots & \frac{(x_{j-1} - x_j)^i}{i!} & \cdots & \frac{(x_{j-1} - x_j)^{N-1}}{(N-1)!} \\
x_{j+1} - x_j & \frac{(x_{j+1} - x_j)^2}{2} & \cdots & \frac{(x_{j+1} - x_j)^i}{i!} & \cdots & \frac{(x_{j+1} - x_j)^{N-1}}{(N-1)!} \\
\cdots & \cdots & \cdots & \cdots & \cdots & \cdots \\
x_N - x_j & \frac{(x_N - x_j)^2}{2} & \cdots & \frac{(x_N - x_j)^i}{i!} & \cdots & \frac{(x_N - x_j)^{N-1}}{(N-1)!}
\end{array}
\right),\label{4.6}
\end{align}
\begin{align}
	\Delta F = \left(\mathcal{Z}(x_1) - \mathcal{Z}(x_j), \mathcal{Z}(x_2) - \mathcal{Z}(x_j), \cdots, \mathcal{Z}(x_{j-1}) - \mathcal{Z}(x_j), \mathcal{Z}(x_{j+1}) - \mathcal{Z}(x_j), \cdots, \mathcal{Z}(x_N) - \mathcal{Z}(x_j)\right)^T,\label{4.7}
\end{align}
\begin{align}
	D = \left(\mathcal{Z}'(x_j), \mathcal{Z}''(x_j), \cdots, \mathcal{Z}^{(n)}(x_j), \cdots \mathcal{Z}^{(N-1)}(x_j)\right)^T.\label{4.8}
\end{align}
\end{widetext}
According to Cramer’s rule, the derivative terms in Eq.\eqref{4.8} can be formally solved from Eq.\eqref{4.5}. In particular, the first and second derivatives are  
\begin{align}
	\mathcal{Z}'(x_j) =\dfrac{\text{det}(M_1)}{\text{det}(M)},~~ \mathcal{Z}''(x_j)=\dfrac{\text{det}(M_2)}{\text{det}(M)}.\label{4.9}
\end{align}
$M_i(i=1,2)$ is a matrix obtained by replacing the $i$-th column of $M$ with $\Delta F$. In Eq.\eqref{4.9}, the derivatives are expressed as linear combinations of function values at the $N$ discrete points.

With Eq.\eqref{4.9}, the differential equation \eqref{4.3}  can be written as a matrix equation,
\begin{align}
\bar{\mathcal{M}} \mathcal{F} = 0,\label{4.10}
\end{align}
where
\begin{align}
\mathcal{F} = \left(\mathcal{Z}(x_1),\mathcal{Z}(x_2), \cdots, \mathcal{Z}(x_N)\right)^T.\label{4.11}
\end{align}
The boundary conditions $\mathcal{Z}(0)=\mathcal{Z}(1)=0$ can be imposed by approximating Eq.\eqref{4.10} as
\begin{align}
\mathcal{M} \mathcal{F} = 0,\label{4.12}
\end{align}
where
\begin{align}
\left(\mathcal{M}\right)_{j,i} = \left\{
\begin{array}{cc}
\delta_{j,i}, & j=1 \text{ or } N, \\
\left(\bar{\mathcal{M}}\right)_{j,i}, & j=2,3,\cdots,N-1
\end{array}
\right.\label{4.13}
\end{align}
In order to have a nonzero solution for $\mathcal{F}$ in Eq.\eqref{4.12}, the determinant of $\mathcal{M}$ should vanish. The QNM frequencies are thus determined by 
\begin{equation}
\text{det}(\mathcal{M})=0.
\end{equation} 

\subsection{Results}

With the methods described previously in this section, we here calculate the fundamental QNM frequencies for different values of the model parameters $\left\lbrace k,m_1,m_2,a,b,\mu \right\rbrace$. 

To validate our numerical codes, we first calculate the QNMs of scalar perturbation in $5D$ Schwarzschild black hole and singly rotating MPBH, and compare them with previous results in literature \cite{Zhidenko:2006rs,Li:2023wog}, which are shown in Table \ref{tab:1.1}.
We find good agreement between our results with continued fraction method ($\omega_{\text{CFM}}$) and matrix method ($\omega_{\text{MM}}$) and the ones in literature ($\omega_{\text{Ref}}$).
\begin{table}[tbph]
	\centering
	\caption{Comparison between QNMs calculated with our methods and ones in literature. $M=1, b=0$.}
	\label{tab:1.1}
	\renewcommand\arraystretch{1.25}
	\setlength{\tabcolsep}{2.5pt}{
		\begin{tabular}{c c c c c}
			\hline 
			\hline
			  $\left\lbrace a,\mu,k,m_1,m_2 \right\rbrace$ & $$ & $\omega_{\text{Ref}}$ & $\omega_{\text{MM}}$ & $\omega_{\text{CFM}}$\\
              \hline
			$\left\lbrace0,0,0,0,0\right\rbrace$ & $\text{Re}$& $0.53384 $ & $0.534406 $ &$0.533838  $\\
            $ $ &$-\text{Im}$&  $  0.38338i$ & $  0.383241 i$ &$  0.383387i $\\
            \hline
            $\left\lbrace0,0,0,1,0\right\rbrace$ & $\text{Re}$& $1.01602 $ & $1.01620 $ &$1.01602  $\\
            $$ &$-\text{Im}$&  $  0.36233i$ & $  0.362074 i$ &$ 0.362328 i $\\	
            \hline
			$\left\lbrace0,0,0,1,1\right\rbrace$ & $\text{Re}$& $1.51057 $ & $1.51055 $ &$1.51057  $\\
            $$ &$-\text{Im}$&  $  0.35754i$ & $  0.357371 i$ &$  0.357537i $\\
            \hline
			$\left\lbrace0.2,0,1,1,1\right\rbrace$ & $\text{Re}$& $ 2.565177 $ & $2.5651 $ & $2.56514 $\\
            $$ & $-\text{Im}$& $   0.353047i$ & $  0.353033 i$ & $  0.353072 i$\\
            \hline
            $\left\lbrace0.5,0,1,1,0\right\rbrace$&$\text{Re}$& $2.178379 $ & $2.17821$ &$2.17828 $\\
             $$&$-\text{Im}$& $  0.340853i$ & $  0.340858 i$ &$  0.340943 i$\\
            \hline
			$\left\lbrace0.3,0.3,1,1,1\right\rbrace$&$\text{Re}$& $2.609174 $ & $2.60908 $ &$2.60913$\\
            $$&$-\text{Im}$& $ 0.348709i$ & $  0.348761 i$ &$  0.348797 i$\\            
			\hline
			\hline
	\end{tabular}}
\end{table}

In Table \ref{tab:2}, we show some examples of the fundamental massive scalar QNMs in MPBH with two unequal rotation parameters. We also find good agreement between the results with our two methods. The results $\omega_{\text{CFM}}$ in this table are calculated with a 25-term recursion relation. 
\begin{table}[tbph]
	\centering
	\caption{Comparison between QNMs with two methods. $M=1, m_1=1, m_2=1$.}
	\label{tab:2}
	\renewcommand\arraystretch{1.25}
	\setlength{\tabcolsep}{2.5pt}{
		\begin{tabular}{c c c c c c}
			\hline 
			\hline
			$a$ & $b$ & $\mu$ & $k$ & $\omega_{\text{MM}}$ & $\omega_{\text{CFM}}$\\
			\hline
			$0.2$ & $0.3$ & $0.1$ & $0$ & $1.6812 - 0.347026 i$ &$1.68112 - 0.3472 i$\\
			$0.4$ & $0.2$ & $0.9$ & $0$ & $1.8222 - 0.311043 i$ & $1.82222 - 0.311152 i$\\
			$0.3$ & $0.1$ & $0.1$ & $1$ & $2.63435 - 0.349055 i$ &$2.6344 - 0.349097 i$\\
			\hline
			\hline
	\end{tabular}}
\end{table}

As mentioned before, the continued fraction method is applicable in the approximate condition $r_2\ll1$. Therefore, we mainly use the matrix method to calculate the QNM frequencies in the rest of the paper.
For the cases where $r_2<0.1$, we will use the continued fraction method with a 25-term recursion relation as a complementary method and to provide crosschecks. 

\begin{figure}[h] 
    \centering 
    \includegraphics[width=0.5\textwidth]{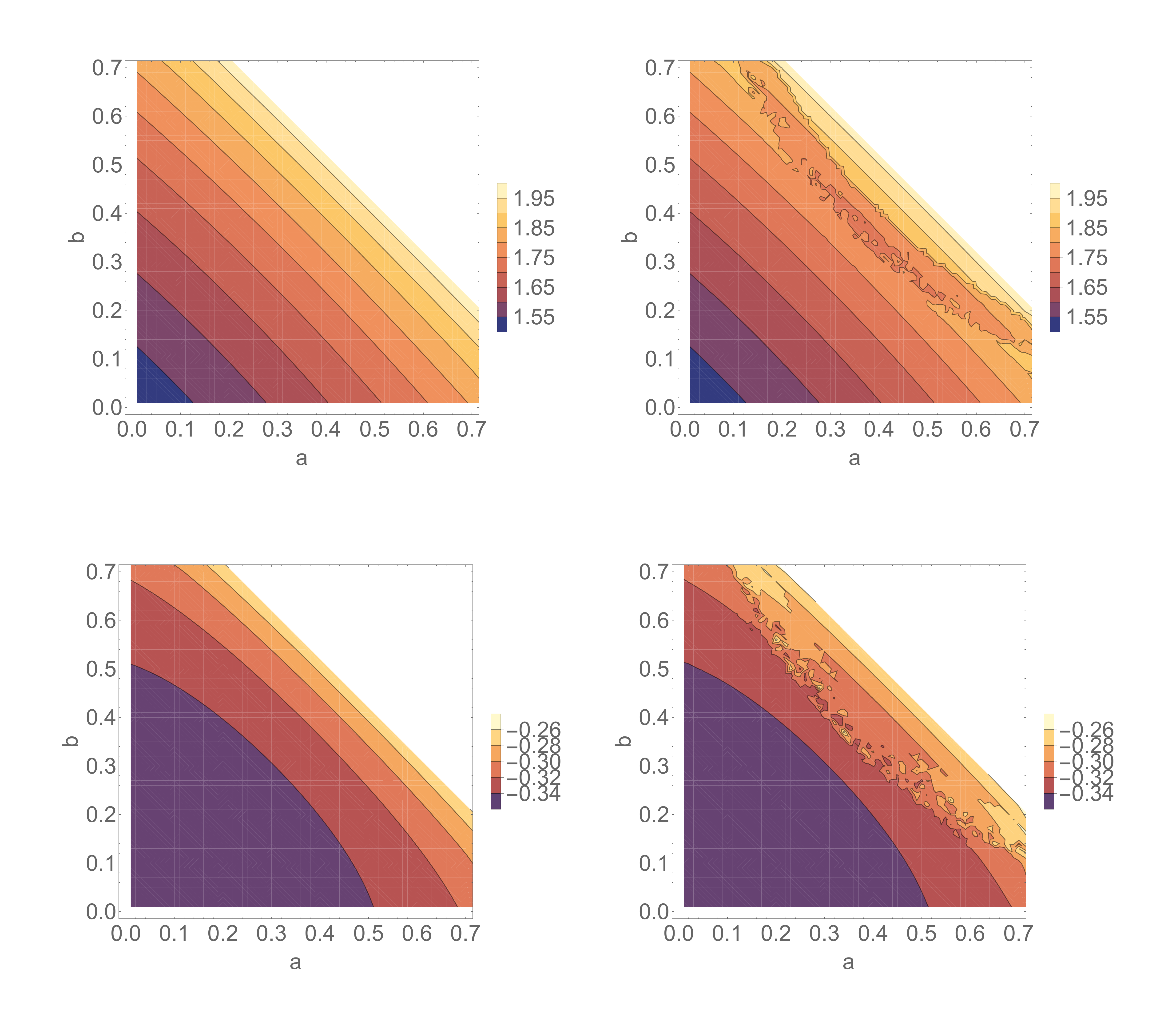}  
    \caption{$\mathrm{Re}(\omega)$ and $\mathrm{Im}(\omega)$ of the QNM frequencies $\omega$ for different values of $a$ and $b$. The step lengths are  
$\Delta a = 0.01, \Delta b = 0.01$. Other parameters are $ M=1,\mu=0.1,k=0,m_1=1,m_2=1$. Panels in top row are $\mathrm{Re}(\omega)$ and in bottom row are $\mathrm{Im}(\omega)$.
Panels in left column are matrix method results and in right column are continued fraction results.
} 
    \label{fig:ab3d} 
\end{figure}

In Fig.\ref{fig:ab3d}, we show the real and imaginary parts of the massive scalar QNMs with different values of $a$ and $b$. The other model parameters are chosen as $\left\lbrace M=1,\mu=0.1,k=0,m_1=1,m_2=1, \right\rbrace$. The first row shows the real parts and the second row shows the imaginary parts. In each row, the left panel is calculated with matrix method, and the right panel is calculated with continued fraction method. 
We can see that the results in right panels begin to show instability when the value of $r_2\gtrsim 0.1$. From the radial and angular EOMs, one can check that there is a symmetry for QNM frequencies $\omega$ under $a\leftrightarrow b$ when $m_1=m_2$. This symmetry is obvious in our numerical results. As $a$ or $b$ increases, the real part of $\omega$ increases, and the imaginary part of $\omega$ is negative with decreasing absolute value. This is the same as the case in $5D$ singly rotating MPBH \cite{Li:2023wog}. All found QNMs are decaying modes. Certain numerical results of the QNMs are listed in Table \ref{tab:011}.   
\begin{table*}[]
	\centering
	 \caption{The fundamental QNMs with $k = 0,m_1 =1, m_2 = 1,\mu=0.1$ for different values of $a$ and $b$.}
	\label{tab:011}
	\setlength{\tabcolsep}{1mm}
	\renewcommand\arraystretch{1.5}
	\begin{tabular}{c|c|c|c|c|c|c}
		\hline 
		\hline
		$b \backslash a$ & $0.1$ & $0.2$ & $0.3$ & $0.4$& $0.5$& $0.6$ \\
		\hline
		$0.1$ & $1.56895 - 0.355839 i$ & $1.60232 - 0.354012 i$ & $1.64036 - 0.350607 i$ & $1.68394 - 0.345147 i$ & $1.73441 - 0.33684 i$&$1.79398 - 0.324253 i$\\
        \hline
		$0.2$ & $1.60232 - 0.354012 i$ & $1.63886 - 0.351547 i$ & $1.6812 - 0.347026 i$ & $1.73079 - 0.339647 i$ & $1.78996 - 0.327913 i$ & $1.86312 - 0.308557 i$\\
        \hline
		$0.3$ & $1.64036 - 0.350607 i$ & $1.6812 - 0.347026 i$ & $1.72958 - 0.340538 i$ & $1.78795 - 0.329626 i$ & $1.86084 - 0.310928 i$ & $1.95865 - 0.274389 i$ \\
        \hline
		$0.4$ & $1.68394 - 0.345147 i$ & $1.73079 - 0.339647 i$ & $1.78795 - 0.329626 i$ & $1.86008 - 0.311689 i$ & $1.95776 - 0.275621 i$ & $-$ \\
        \hline
		$0.5$ & $1.73441 - 0.33684 i$ & $1.78996 - 0.327913 i$ & $1.86084 - 0.310928 i$ & $1.95776 - 0.275621 i$ & $-$ & $-$\\
        \hline
		$0.6$ & $1.79398 - 0.324253 i$ & $1.86312 - 0.308557 i$ & $1.95865 - 0.274389 i$ & $-$ & $-$ & $-$\\
		\hline
		\hline
	\end{tabular}
\end{table*}

\begin{figure}[h] 
    \centering 
    \includegraphics[width=0.4\textwidth]{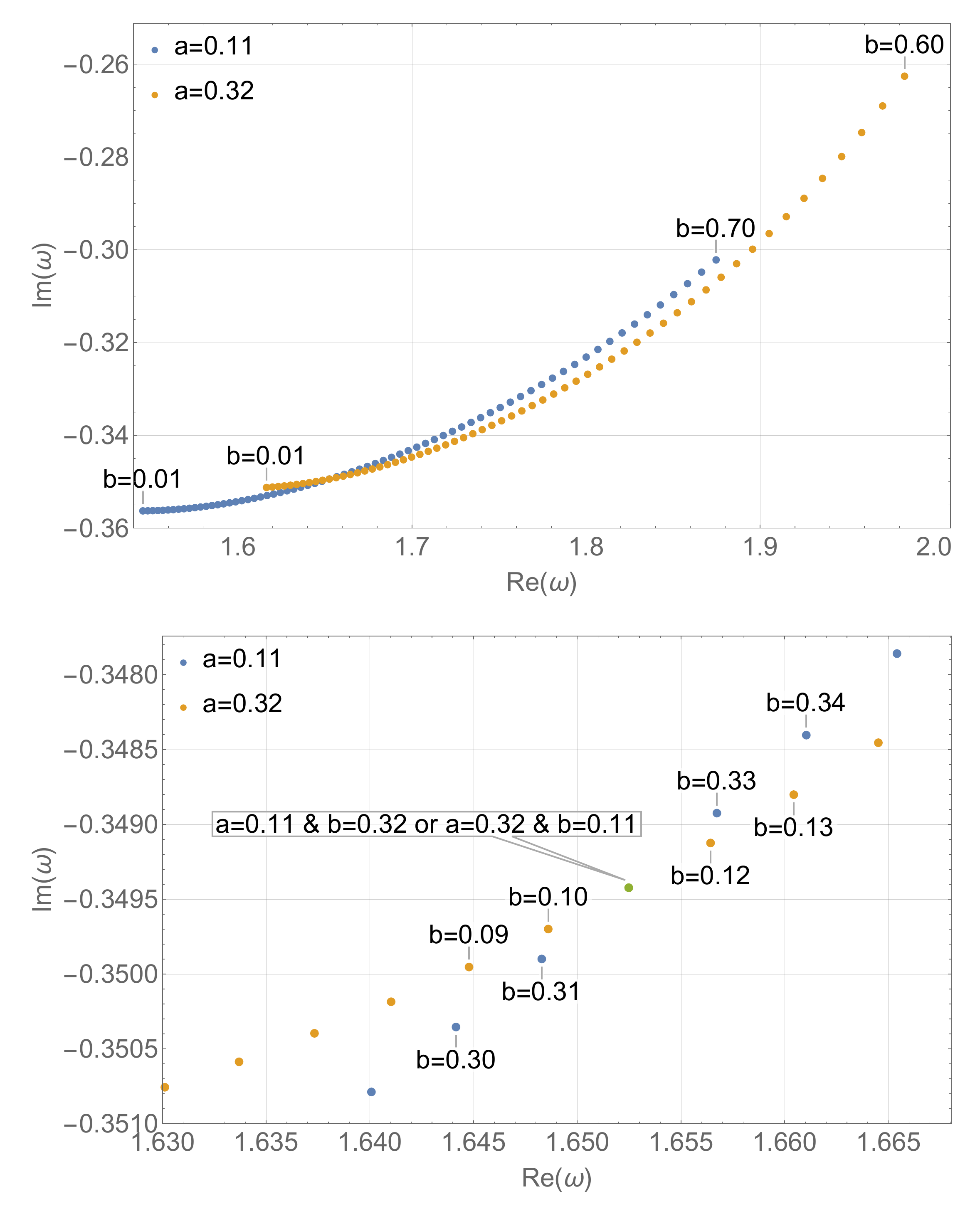}  
    \caption{Impact of the rotation parameter $b$ on $\mathrm{Re}(\omega)$ and $\mathrm{Im}(\omega)$ of the QNM frequency. The step length 
$\Delta b = 0.01$. The intersection point (in Green) demonstrates the symmetry under $a\leftrightarrow b$. Other parameters are $ M=1,\mu=0.1,k=0,m_1=m_2=1$.} 
    \label{fig:ab} 
\end{figure}

In Fig.\ref{fig:ab}, the top panel is the variation of the QNMs with $b$ on the complex $\omega$-plane and the bottom panel is the zoom-in around the intersection point in the top panel. In the top panel, it is obvious that the real part of QNM frequency increase as $b$ increases while the absolute value of the imaginary part decreases. The larger the rotation parameter is, the longer the scalar quasinormal ringing lifetime is. In the bottom panel,
the Green intersection point of the blue and orange dotted line explicitly demonstrates the symmetry of QNMs under $a\leftrightarrow b$ in current parameter setting.
\begin{figure}[h] 
    \centering 
    \includegraphics[width=0.43\textwidth]{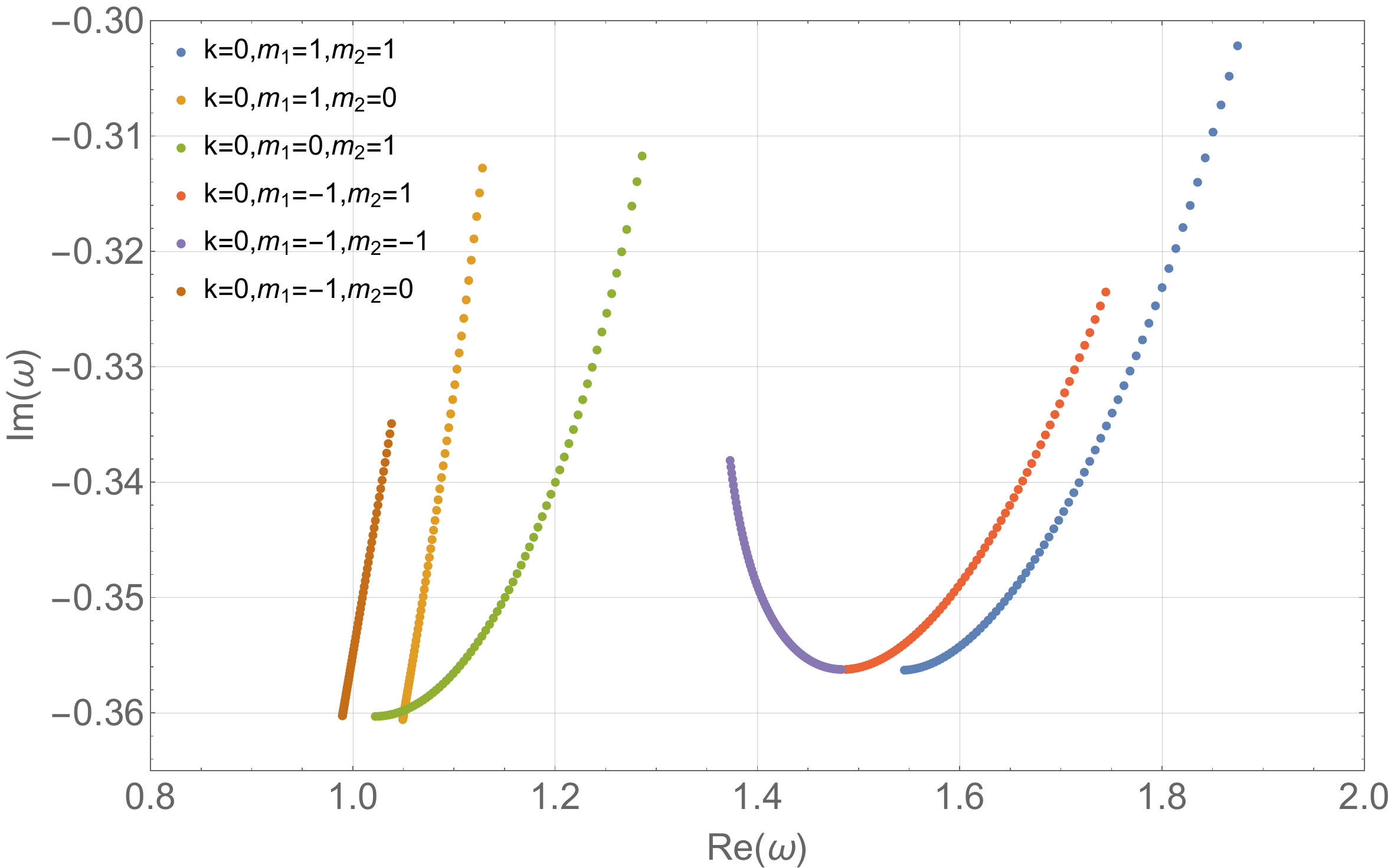}  
    \caption{The variation of the fundamental QNM frequency with parameter $b$ for cases with different azimuthal numbers. In each dotted line, the lowest point corresponds to $b=0$ and the step length is $\Delta b = 0.01$. Other parameters are $ M=1,\mu=0.1,a=0.11$.} 
    \label{fig:km1m2} 
\end{figure}

\begin{figure}[h] 
    \centering 
    \includegraphics[width=0.43\textwidth]{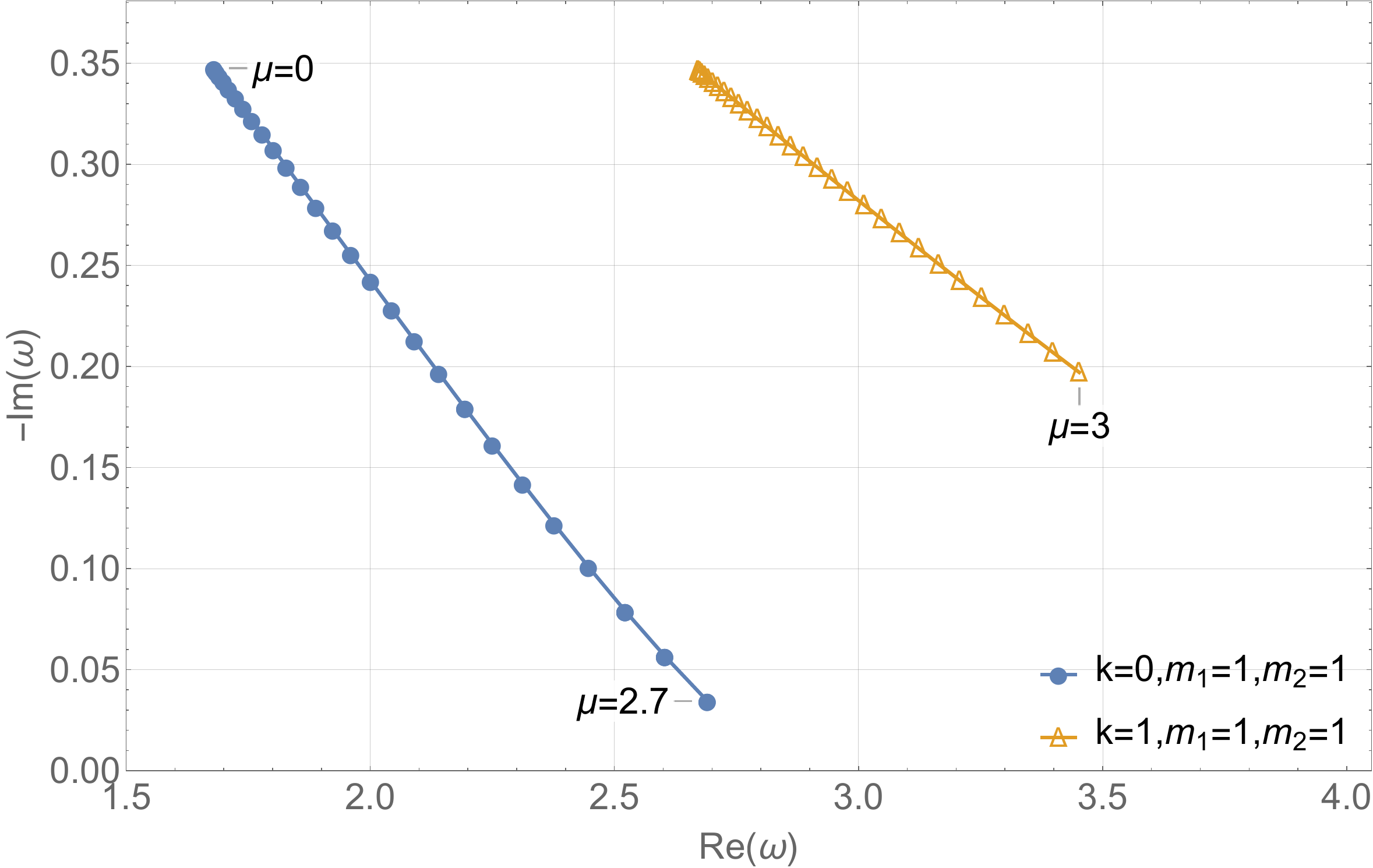}  
    \caption{The values of $\mathrm{Re}(\omega)$ and $-\mathrm{Im}(\omega)$ of the QNM frequencies for different $\mu$. The step length is
$\Delta \mu = 0.1$. Other parameters are $ M=1,a=0.2,b=0.3.$} 
    \label{fig:mu} 
\end{figure}

In Fig.\ref{fig:km1m2}, we consider the impact of azimuthal numbers $m_1,m_2$ on the variation of the QNM frequency with the rotation parameter $b$. In each dotted line, $b=0$ for the lowest point and $b=0.7$ for the highest one. We can see that the absolute value of the imaginary part of the QNM frequency decreases as $b$ increases in each case. 
The real part of corotating ($m_2>0$) QNM frequency increases as $b$ increases while the real part of counterrotating ($m_2<0$) QNM frequency decreases. This is similar as the four-dimensional Kerr black hole case \cite{Dolan:2007mj}.

In Fig.\ref{fig:mu}, we plot the values of $\mathrm{Re}(\omega)$ and $-\mathrm{Im}(\omega)$ of the QNM frequencies for different $\mu$. 
It is easy to notice that the imaginary parts of the QNMs tend to zero as the scalar mass $\mu$ increases. These long-living modes are qualitatively 
the same as that found in Kerr black hole cases \cite{Konoplya:2006br,Dolan:2007mj}, Schwarzschild cases \cite{Ohashi:2004wr,Konoplya:2004wg,Zhidenko:2006rs} and singly rotating $5D$ MPBH case \cite{Li:2023wog}.
It is also found that the imaginary part $\mathrm{Im}(\omega)$  has a more slower tendency to zero for larger $k$.

In Tables \ref{tab:110}, \ref{tab:101} and \ref{tab:111}, we list certain values of the QNMs for different $a$ and $b$. Different tables correspond to different choices of  $\{k ,m_1, m_2\}$. Since $m_1\neq m_2$, the symmetry of QNMs under $a\leftrightarrow b$ disappears in Table \ref{tab:110} and Table \ref{tab:101}. 
However, by comparing Table \ref{tab:110} and Table \ref{tab:101}, we can see another symmetry: when we exchange not only the values of $a$ and $b$, but also $m_1$ and $m_2$, the QNM frequencies is invariant.

It is known that general $5D$ MPBH has $U(1)\times U(1)$ symmetry while the degenerate MPBH ($a=b$) has enhanced $U(2)$ symmetry \cite{Murata:2007gv,Frolov:2002xf}. 
The enhanced symmetry leads to the fact that the QNM frequencies don't depend on parameter $k$. We can check this fact by comparing results in Table \ref{tab:011} and Table \ref{tab:111}. Due to the different symmetries for cases with $a=b$ and $a\neq b$, from Tables \ref{tab:110}, \ref{tab:101}, \ref{tab:111}, we can also see that the real parts of the QNMs with $k>0$ have sharp decreases while $a=b$. 
\begin{table*}[]
	\centering
	 \caption{The fundamental QNMs with $k = 1,m_1 =1, m_2 = 0,\mu=0.1$ for different values of $a$ and $b$.}
	\label{tab:110}
	\setlength{\tabcolsep}{1mm}
	\renewcommand\arraystretch{1.5}
	\begin{tabular}{c|c|c|c|c|c|c}
		\hline 
		\hline
		$b \backslash a$ & $0.1$ & $0.2$ & $0.3$ & $0.4$& $0.5$& $0.6$ \\
		\hline
		$0.1$ & $1.04769 - 0.360006 i$ & $2.06828 - 0.352844 i$ & $2.10233 - 0.349991 i$ & $2.14023 - 0.345594 i$ & $2.18235 - 0.339247 i$ & $2.22911 - 0.33035 i$\\
        \hline
		$0.2$ & $2.04284 - 0.352826 i$ & $1.08355 - 0.35586 i$ & $2.10912 - 0.347573 i$ & $2.14826 - 0.342413 i$ & $2.192 - 0.334849 i$ & $2.24086 - 0.323904 i$\\
        \hline
		$0.3$ & $2.05147 - 0.350037 i$ & $2.08406 - 0.34753 i$ & $1.12891 - 0.345735 i$ & $2.16212 - 0.336585 i$ & $2.20875 - 0.326514 i$ & $2.26124 - 0.311063 i$ \\
        \hline
		$0.4$ & $2.06391 - 0.34587 i$ & $2.09849 - 0.342359 i$ & $2.1378 - 0.336449 i$ & $1.19003 - 0.321012 i$ & $2.23358 - 0.312024 i$ & $-$ \\
        \hline
		$0.5$ & $2.08056 - 0.339985 i$ & $2.11801 - 0.334817 i$ & $2.16113 - 0.326066 i$ & $2.21081 - 0.311453 i$ & $-$ & $-$\\
        \hline
		$0.6$ & $2.10205 - 0.331786 i$ & $2.14357 - 0.323765 i$ & $2.19196 - 0.309667 i$ & $-$ & $-$ & $-$\\
		\hline
		\hline
	\end{tabular}
\end{table*}
\begin{table*}[]
	\centering
	 \caption{The fundamental QNMs with $k = 1,m_1 =0, m_2 = 1,\mu=0.1$ for different values of $a$ and $b$.}
	\label{tab:101}
	\setlength{\tabcolsep}{1mm}
	\renewcommand\arraystretch{1.5}
	\begin{tabular}{c|c|c|c|c|c|c}
		\hline 
		\hline
        $b \backslash a$ & $0.1$ & $0.2$ & $0.3$ & $0.4$ & $0.5$ & $0.6$ \\
        \hline
        $0.1$ & $1.04769 - 0.360006 i$ & $2.04284 - 0.352826 i$ & $2.05147 - 0.350037 i$ & $2.06391 - 0.34587 i$ & $2.08056 - 0.339985 i$ & $2.10205 - 0.331786 i$ \\
        \hline
        $0.2$ & $2.06828 - 0.352844 i$ & $1.08355 - 0.35586 i$ & $2.08406 - 0.34753 i$ & $2.09849 - 0.342359 i$ & $2.11801 - 0.334817 i$ & $2.14357 - 0.323765 i$ \\
        \hline
        $0.3$ & $2.10233 - 0.349991 i$ & $2.10912 - 0.347573 i$ & $1.12891 - 0.345735 i$ & $2.1378 - 0.336449 i$ & $2.16113 - 0.326066 i$ & $2.19196 - 0.309667 i$ \\
        \hline
        $0.4$ & $2.14023 - 0.345594 i$ & $2.14826 - 0.342413 i$ & $2.16212 - 0.336585 i$ & $1.19003 - 0.321012 i$ & $2.21081 - 0.311453 i$ & $-$ \\
        \hline
        $0.5$ & $2.18235 - 0.339247 i$ & $2.192 - 0.334849 i$ & $2.20875 - 0.326514 i$ & $2.23358 - 0.312024 i$ & $-$ & $-$ \\
        \hline
        $0.6$ & $2.22911 - 0.33035 i$ & $2.24086 - 0.323904 i$ & $2.26124 - 0.311063 i$ & $-$ & $-$ & $-$ \\	
		\hline
		\hline
	\end{tabular}
\end{table*}

\begin{table*}[]
	\centering
	 \caption{The fundamental QNMs with $k = m_1 = m_2 = 1,\mu=0.1$ for different values of $a$ and $b$ .}
	\label{tab:111}
	\setlength{\tabcolsep}{1mm}
	\renewcommand\arraystretch{1.5}
	\begin{tabular}{c|c|c|c|c|c|c}
		\hline 
		\hline
		$b \backslash a$ & $0.1$ & $0.2$ & $0.3$ & $0.4$ & $0.5$ & $0.6$ \\
        \hline
        $0.1$ & $1.56895 - 0.355839 i$ & $2.59614 - 0.352072 i$ & $2.63435 - 0.349055 i$ & $2.67832 - 0.344386 i$ & $2.72896 - 0.337559 i$ & $2.78765 - 0.327711 i$ \\
        \hline
        $0.2$ & $2.59614 - 0.352072 i$ & $1.63886 - 0.351547 i$ & $2.67298 - 0.346169 i$ & $2.72134 - 0.340239 i$ & $2.77812 - 0.33127 i$ & $2.84569 - 0.317518 i$ \\
        \hline
        $0.3$ & $2.63435 - 0.349055 i$ & $2.67298 - 0.346169 i$ & $1.72958 - 0.340538 i$ & $2.77349 - 0.332904 i$ & $2.83939 - 0.31984 i$ & $2.92089 - 0.297536 i$ \\
        \hline
        $0.4$ & $2.67832 - 0.344386 i$ & $2.72134 - 0.340239 i$ & $2.77349 - 0.332904 i$ & $1.86008 - 0.311689 i$ & $2.91735 - 0.298797 i$ & $-$ \\
        \hline
        $0.5$ & $2.72896 - 0.337559 i$ & $2.77812 - 0.33127 i$ & $2.83939 - 0.31984 i$ & $2.91735 - 0.298797 i$ & $-$ & $-$ \\
        \hline
        $0.6$ & $2.78765 - 0.327711 i$ & $2.84569 - 0.317518 i$ & $2.92089 - 0.297536 i$ & $-$ & $-$ & $-$ \\
		\hline
		\hline
	\end{tabular}
\end{table*}

\section{Summary}

In this paper, we study the massive scalar QNMs of a $5D$ doubly rotating MPBH with two arbitrary rotation parameters. Two numerical methods are used, namely, the continued fraction method and matrix method. We make extensive calculation for the fundamental QNMs with various model parameters $\{a,b,\mu,k,m_1,m_2\}$, and show our numerical results in several tables and figures.

It is found that all the obtained fundamental QNMs are decaying modes. The absolute values of the imaginary parts of the QNMs always decrease as the rotation parameters increase. 
As the scalar mass $\mu$ increases, the imaginary parts of the QNMs have a tendency to zero, which results in the existence of long-living modes.

 As the rotation parameters increase, it is found that the real parts of the corotating QNMs with $k=0$ increase while the real parts of the counterrotating QNMs with $k=0$ decrease. The corotating QNMs with higher azimuthal numbers vary more rapidly with rotation parameters.    

From the radial and angular EOMs, we know the QNMs are invariant under $(a,m_1)\leftrightarrow(b,m_2)$ or in particular, $a\leftrightarrow b$ when $m_1=m_2$. The calculated numerical results in the tables and figures demonstrate these symmetries explicitly.    

From our numerical results, we also see that the real parts of the QNMs with $k>0$ have a big depression while $a=b$, which hits the enhanced symmetry of the degenerate MPBH.

\section*{Acknowledgments}
The authors would like to thank Zhan-Feng Mai, Ya-Ting Peng, Jinbo Yang and Ai-Xu Zhou for their valuable discussions. This work is partially supported by Guangdong Major Project of Basic and Applied Basic Research (No.2020B0301030008).

\end{document}